\documentstyle[A4,11pt,epsfig,twoside]{article}
\def\ma{m_{\chi_a}}
\def\mb{m_{\chi_b}}
\def\etal{{\it et al.}}
\begin{document}
\setcounter{page}{1}
\title{CHARGINO PAIR PRODUCTION AT ONE-LOOP}
\author{H. BAER${}^1$, M.A. D\'IAZ${}^2$\thanks{Speaker.
Presented at Workshop on Physics and Experiments with Future 
Electron-Positron Linear Colliders August 26-30, 2002, Jeju Island, Korea.
}\,\,, 
M.A. RIVERA${}^2$, AND D.A. ROSS${}^3$
\\
\\
{\it ${}^1$Dept. of Physics, Florida State University, Tallahassee FL 32306,
USA}
\\
{\it ${}^2$D. de F\'\i sica, U. Cat\'olica de Chile, Av. V. Mackenna 
4860, Santiago, Chile}
\\
{\it ${}^3$Dept. of Physics, U. of Southampton, Southampton SO17 1BJ, U.K.}
}
\date{}
\maketitle
\begin{abstract}
We use the quantum corrected chargino production cross section, including
information on the beam polarization and chargino helicities, to estimate 
the precision achievable at a 1 TeV, 500 ${\mathrm fb}^{-1}$ Linear 
Collider on the determination of fundamental supersymmetric parameters
from measurements of light chargino mass and production cross sections.
We show that to get meaningful results higher order corrections should be 
included. 
\end{abstract}


It has been established that it will be possible to make precision 
measurements of supersymmetric observables at a future Linear 
Collider \cite{LC}. Therefore, it is imperative to match the experimental 
precision with higher order theoretical calculations \cite{Review}.
Chargino production at an LC,
\begin{equation}
e^+(p_2) + e^-(p_1) \quad \longrightarrow \quad 
\tilde\chi^+_b(k_2) + \tilde\chi^-_a(k_1)
\end{equation}
is a well studied laboratory for the determination of the parameters of the 
MSSM. The total unpolarized cross section was first calculated at one-loop in
the approximation where only quarks and squarks were included \cite{qsq},
while in the same approximation, the cross section with polarized beams 
was found in \cite{DKR2}. These early works showed the importance of the 
quark-squark loops. The first complete one-loop calculation of the 
polarized production cross section is in \cite{BH}, where the importance 
of box graphs was demonstrated. This was followed by the first calculation 
of polarized cross sections which keeps the full information on the 
helicity state of the charginos and includes self energies, triangles and 
boxes, excluding only the QED corrections \cite{prototype,DRhelicity}. 
Here the cross sections are given in terms of helicity amplitudes,
\begin{equation}
\frac{d\sigma(\alpha,\lambda_2,\lambda_1)}{d\cos\theta}
  \ = \ \frac{\lambda^{1/2}(s,\ma^2,\mb^2)}{128 \, \pi \, s} 
 \left| {\cal A}^\alpha_{\lambda_2,\lambda_1} \right|^2, 
\end{equation}
where $\lambda_1$ and $\lambda_2$ are the helicities of the $\chi^-$ and 
$\chi^+$ respectively, and $\alpha$ is the electron polarization. The 
renormalized helicity amplitudes are given in terms of generalized 
$Q$-charges, {\sl e.g.},
\begin{eqnarray}
{\cal A}^L_{+-}&=&
-{\cal Q}^L_{31}(1+v)\,(1+\cos\theta)
-{\cal Q}^L_{32}\,s\,v\,(1+v)\,\sin^2\theta
\nonumber\\ &&
-{\cal Q}^L_{41}(1-v)\,(1+\cos\theta)
-{\cal Q}^L_{42}\,s\,v\,(1-v)\,\sin^2\theta
\nonumber\\ &&
-4{\cal Q}^L_{52}\sqrt{s}\,\sqrt{1 - v^2}\,(1+\cos\theta)
\end{eqnarray}
where $v$ is the velocity of the chargino and $\theta$ is the scattering 
angle. All quantum corrections are included in the $Q$-charges.

Here we are interested in the determination of MSSM parameters from chargino
observables hypothetically taken at an LC with 1 TeV center of mass energy
and $500\, {\mathrm fb}^{-1}$ integrated luminosity. To this end we use
a $\chi^2$ analysis whose input is the Supergravity benchmark model 
B \cite{benchmark} with $m_0=100$ GeV, $M_{1/2}=250$ GeV, $\tan\beta=10$, 
and $\mu>0$. In this model, the lightest chargino has a mass
$m_{\chi^+_1}=181$ GeV and it is pair produced at the LC with an unpolarized
cross section $\sigma=420$ fb, expecting of the order of $10^5$ events.
Other observables we include in our $\chi^2$ are the left cross sections
$\sigma_L^{++}=15$ fb, $\sigma_L^{+-}=250$ fb, and $\sigma_L^{-+}=140$ fb, 
and the right cross sections $\sigma_R^{++}=0.05$ fb, $\sigma_R^{+-}=0.44$ 
fb, and $\sigma_R^{-+}=0.29$ fb. We do not include observables from the heavy
chargino. For the cross sections we consider only the statistical error
($N\pm\sqrt{N}$), and for the chargino mass we take a 1\% error.

\begin{figure}
\centerline{\protect\hbox{\epsfig{file=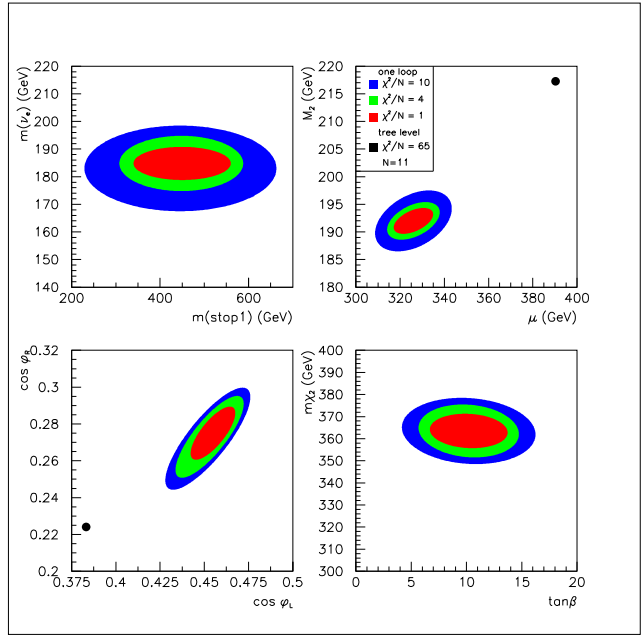,width=0.6\textwidth}}}
\vskip .5cm
\caption[]{
Regions of parameter space where normalized $\chi^2\le 1$ (inner ellipse). 
For comparison, also shown are the $\chi^2\le 4$ (central) and $\chi^2<10$ 
(outer) regions.
}
\label{chi_elipse}
\end{figure}
The results of the $\chi^2$ analysis are summarized in 
Fig.~\ref{chi_elipse}, where regions of normalized $\chi^2<1$ are depicted
(inner ellipses). For comparison we also show the regions where $\chi^2<4$
(central ellipses) and $\chi^2<10$ (outer ellipses). The center of each 
ellipse indicates the output of the $\chi^2$ analysis which should coincide 
with the input, and the width of each ellipse indicates the error in the 
determination of the corresponding parameter. Filled circles in two of the 
quadrants correspond to the tree level prediction, with a mediocre 
normalized $\chi^2=65$. These results are summarized in Table 
\ref{tab:cases}.

\begin{table}
\begin{center}
\caption{
Parameter determination from light chargino pair production at one-loop.
All dimensionful parameters are expressed in GeV.
}
\bigskip
\begin{tabular}{lccccc}
\hline
parameter & input & output  & error & percent  \\
\hline
$M_2             $ & 193.6 & 192  & 3    & 2  \\
$\mu             $ & 328.2 & 326  & 9    & 3  \\
$\tan\beta       $ & 10.0  & 10   & 3    & 30 \\
$m_{\tilde\nu_e} $ & 188.0 & 185  & 6    & 3  \\
$\cos\phi_L      $ & 0.452 & 0.46 & 0.01 & 2  \\
$\cos\phi_R      $ & 0.273 & 0.28 & 0.01 & 4  \\
$m_{\chi^+_2}    $ & 365.0 & 364  & 8    & 2  \\
$m_{\tilde t_1}  $ & 392.0 & 450  & 110  & 24 \\
\hline
\label{tab:cases}
\end{tabular}
\end{center}
\end{table}
From the parameters in the chargino mass matrix at tree level, the gaugino
mass $M_2$ and the higgsino mass $\mu$ can be determined within 2-3\%, but 
$\tan\beta$ is determined within $30\%$ due to a weaker dependence of the
observables on this parameter. The tree level mass matrix is diagonalized
with two rotation matrices defined by the angles $\phi_L$ and $\phi_R$
and their cosine can be found with a 2-4\% error. The heavy chargino and
the sneutrino masses (the former intervenes at tree level into the 
production cross section with left handed polarized electrons) can be found 
with a 2-3\% error. Last but not least, we can see in Fig.~\ref{chi_elipse}
the determination of the stop quark mass $m_{\tilde t_1}$, which is a pure
one-loop effect: it is found to be $m_{\tilde t_1}=450\pm110$ GeV, while the
input from benchmark B is $m_{\tilde t_1}=392$ GeV. This is very interesting 
since it indicates that precision measurements in the chargino sector may 
shed light into the squark sector, in the same way that the SM precision 
measurements gave us information on the top quark mass (and after its 
discovery, on the Higgs mass).

In summary, we have shown that in order to determine the underlying 
parameters of the MSSM in the chargino sector, it is necessary to work with 
the one-loop corrected cross sections and masses. In addition, precision 
measurements in the chargino sector can give us information on the squark 
masses.


\end{document}